\documentclass[12pt]{iopart}
\usepackage{graphicx}
\usepackage{array}
\def\lsim{\raise0.3ex\hbox{$\;<$\kern-0.75em\raise-1.1ex
\hbox{$\sim\;$}}}
\def\gsim{\raise0.3ex\hbox{$\;>$\kern-0.75em\raise-1.1ex
\hbox{$\sim\;$}}}
\newcommand{\be}{\begin{equation}}
\newcommand{\ee}{\end{equation}}
\newcommand{\bea}{\begin{eqnarray}}
\newcommand{\eea}{\end{eqnarray}}
 \newcommand{\fh}{f_h^{eq}}
 \newcommand{\fht}{f_{\tilde{h}}^{eq}}
\newcommand{\fLeq}{f_{\ell}^{eq}} 
\newcommand{\fLteq}{f_{\widetilde{\ell}}^{eq}}

\DeclareMathAlphabet{\mathsc}{OT1}{cmr}{m}{sc}
\vbox{%
\hbox{\bf YITP-SB-08-028}}
\begin{document}
\vspace*{.25in}
\title{On Quantum Effects in Soft Leptogenesis}
\author{Chee Sheng Fong} 
\ead{fong@insti.physics.sunysb.edu}
\address{C.N. Yang Institute for Theoretical Physics\\
Stony Brook University, Stony Brook, NY 11794-3840, USA,}
\author{M.C.~Gonzalez-Garcia}
\ead{concha@insti.physics.sunysb.edu}
\address{
C.N.~Yang Institute for Theoretical Physics,\\ 
Stony Brook University, Stony Brook, NY 11794-3840, USA}
\address{and\\
Instituci\'o Catalana de Recerca i Estudis Avan\c{c}ats (ICREA),\\ 
Departament d'Estructura i Constituents de la Mat\`eria,\\
Universitat de Barcelona, 647 Diagonal, E-08028 Barcelona, Spain}
\vspace*{.25in}
\begin{abstract}
It has been recently shown that quantum  Boltzmann equations may 
be relevant for leptogenesis. Quantum effects, which lead to a time-dependent
CP asymmetry, have been shown to be particularly important for 
resonant leptogenesis when the asymmetry is generated by the decay
of two nearly degenerate states. In this work we investigate the
impact of the use of quantum Boltzmann equations in the framework
``soft leptogenesis'' in which supersymmetry soft-breaking terms
give a small mass splitting between the CP-even and CP-odd right-handed
sneutrino states of a single generation and provide the CP-violating phase
to generate the lepton asymmetry. 
\end{abstract}

\maketitle
\section{Introduction}
The discovery of neutrino oscillations makes leptogenesis a very
attractive solution to the baryon asymmetry problem \cite{fy,leptoreview}.  
In the {\sl standard} type I seesaw  framework \cite{ss}, the   
singlet heavy neutrinos have  lepton number violating 
Majorana masses and when decay out of equilibrium produce 
dynamically a lepton asymmetry which  is partially converted into a
baryon asymmetry due to fast sphaleron processes.

For a hierarchical spectrum of right-handed (RH) neutrinos,  
successful leptogenesis requires generically quite heavy singlet neutrino 
masses~\cite{di}, of order $M>2.4 (0.4)\times 10^9$~GeV for vanishing
(thermal) initial neutrino densities~\cite{di,Mbound} 
(although flavour effects \cite{flavour1,flavour2,db2,oscar} 
and/or extended scenarios \cite{db1,ma} may affect this limit).
Low-energy supersymmetry can be invoked to naturally 
stabilize  the hierarchy between this new scale and the 
electroweak one. This, however, introduces a certain conflict between 
the gravitino bound on the reheat temperature and the thermal 
production of RH neutrinos \cite{gravi}. 
A way out of this conflict is provided by resonant 
leptogenesis~\cite{PU1,PU,west}. 
In this scenario RH neutrinos are nearly degenerate in mass 
which makes the 
self energy contributions to the  CP asymmetries resonantly enhanced 
and allowing leptogenesis to be possible at much lower temperatures.

Once supersymmetry has been introduced, leptogenesis is induced also
in singlet sneutrino decays.  If supersymmetry is not broken, the
order of magnitude of the asymmetry and the basic mechanism are the
same as in the non-supersymmetric case. However, as shown in 
Refs.\cite{soft1,soft2}, supersymmetry-breaking terms can play an 
important role in the lepton asymmetry generated in sneutrino decays
because they induce effects which are essentially different 
from the neutrino ones.  In brief, soft supersymmetry-breaking terms
involving the singlet sneutrinos remove the mass degeneracy between
the two real sneutrino states of a single neutrino generation, and
provide new sources of lepton number and CP violation. As a
consequence, the mixing between the sneutrino states generates a
CP asymmetry in their decays. At zero temperature and at lowest order
in the soft supersymmetry-breaking  couplings, the asymmetries 
generated in the sneutrino decays into fermions and scalars cancel out.
However, thermal effects break this cancellation and once they 
are included the asymmetry can be sizable.
In particular it is large for a RH
neutrino mass scale relatively low, in the range $10^{5}-10^{8}$ GeV,
well below the reheat temperature limits, what solves the cosmological
gravitino problem.  
This scenario has been termed ``soft leptogenesis'', since 
the soft terms and not flavour physics provide the necessary mass 
splitting and CP-violating phase.  

In general, soft leptogenesis induced by CP violation in mixing
as discussed above has the drawback that in order to generate
enough asymmetry the lepton-violating  soft bilinear coupling,
responsible for the sneutrino mass splitting, has to be
unconventionally small~\cite{soft1,soft2,ourflasoft}. 
In this
case, as for the case of resonant leptogenesis, the sneutrino self energy 
contributions to the  CP asymmetries are resonantly enhanced
\footnote{
Considering the possibility of CP violation also in decay and in the 
interference of mixing and decay of the sneutrinos~\cite{soft3}, as well 
as extended scenarios~\cite{ourinvsoft,softothers} may alleviate the
unconventionally-small-$B$ problem.}

Till recently, the dynamics of thermal leptogenesis (both 
for the standard see-saw case, as well as for the soft leptogenesis
scenario) has been studied using  the  approach of classical Boltzmann 
equations (BE). The possibility of using quantum Boltzmann equations
(QBE) was first discussed in Ref.~\cite{qbebuch} and it has been recently 
derived in detail in Ref.~\cite{riottoqbe1}. 
In  Ref.~\cite{riottoqbe1}  QBE were obtained 
starting from  the  non-equilibrium quantum field theory based on the 
Closed Time-Path   formulation. They differ from the classical
BE in that  they contain integrals over 
the past times unlike 
in  the classical kinetic theory in which the scattering terms 
do not include any integral over the past history of the 
system which is equivalent to assume that any collision in the plasma  
does not depend upon the previous ones. Quantitatively the most 
important consequence is that  the CP asymmetry 
acquires an additional time-dependent piece,  
with its value at a given instant depending  upon the previous history 
of the system.  If the time variation of the CP asymmetry 
is shorter than the
relaxation time of the particles abundances, the solutions to the
quantum and the classical
Boltzmann equations are expected to differ only by terms of the order
of the ratio of the time-scale of the CP asymmetry to the relaxation
time-scale of the distribution. This is typically the case
in thermal leptogenesis with
hierarchical RH neutrinos . However, as
discussed in Refs.~\cite{riottoqbe2,riottoqbe3}, in the resonant leptogenesis
scenario, $(M_j-M_i)$ is of the order of the decay rate of the RH
neutrinos.  As a consequence the typical time-scale to build up
coherently the time-dependent CP asymmetry, which is of the order of
$(M_j-M_i)^{-1}$, can be larger than the time-scale for the change of 
the abundance of the RH neutrinos. This, as shown in 
Refs.~\cite{riottoqbe2,riottoqbe3}, leads to quantitative 
differences between the classical and the quantum approach in the case of 
resonant leptogenesis and, in particular, in the weak washout
regime they enhance the produced  asymmetry.  
 
Motivated by these results and the fact that in soft leptogenesis the
CP asymmetry is produced resonantly, we perform a detailed study of
the role of quantum effects in the soft leptogenesis scenario.  Our
results show that because of the thermal nature of soft leptogenesis,
the dependence of the quantum effects on the washout regime
for soft leptogenesis is quantitatively different than in the see-saw
resonant scenario.  In particular in the weak washout regime quantum
effects do not enhance but suppress the produced baryon asymmetry.
Quantum effects are most quantitatively important for extremely
degenerate sneutrinos (that is far away from the resonant condition),
$\Delta M \ll \Gamma_{\widetilde N}$, and in the strong washout regime
they can lead to an enhancement of the produced asymmetry,  as well as
change of sign, of the produced asymmetry. But altogether, for a given
$M$ the required values of the lepton violating soft bilinear term $B$
to achieve successful leptogenesis are not substantially modified.

\section{Brief Summary of Soft Leptogenesis: BE and CP asymmetry}
The supersymmetric see-saw model could be described by the superpotential:
\begin{equation}
W=\frac{1}{2}M_{ij}N_{i}N_{j}+Y_{ij}
\epsilon_{\alpha\beta}N_{i}L_{j}^{\alpha}H^{\beta},
\label{eq:superpotential}
\end{equation}
where $i,j=1,2,3$ are flavour indices and $N_{i}$, $L_{i}$, $H$
are the chiral superfields for the RH neutrinos, the left-handed (LH)
lepton doublets and the Higgs doublets with 
$\epsilon_{\alpha\beta}=-\epsilon_{\beta\alpha}$
and $\epsilon_{12}=+1$. 
The corresponding soft breaking terms involving the RH sneutrinos 
$\tilde{N_{i}}$ are given by:
\begin{equation}
\mathcal{L}_{soft}=-\tilde{m}_{ij}^2\widetilde{N}_{i}^{*}\widetilde{N}_{j}
-\left(A_{ij}Y_{ij}\epsilon_{\alpha\beta}\widetilde{N}_{i}
\tilde{\ell}_{j}^{\alpha}h^{\beta}+\frac{1}{2}B_{ij}M_{ij}
\widetilde{N}_{i}\widetilde{N}_{j}+\mbox{h.c.}\right),
\label{eq:soft_terms}
\end{equation}
where $\tilde{\ell}_{i}^{T}=\left(\tilde{\nu}_{i},\tilde{\ell}_{i}^{-}\right)$
and $h^{T}=\left(h^{+},h^{0}\right)$ are the slepton and up-type
Higgs doublets.


As a consequence of the soft breaking $B$ terms, 
the sneutrino and antisneutrino states mix with mass eigenvectors
\begin{eqnarray}
\widetilde{N}_{+i} & = &
\frac{1}{\sqrt{2}}(e^{i\Phi/2}\widetilde{N}_{i}+e^{-i\Phi/2}
\widetilde{N}_{i}^{*}),\nonumber
\\ \widetilde{N}_{-i} & = &
\frac{-i}{\sqrt{2}}(e^{i\Phi/2}
\widetilde{N}_{i}-e^{-i\Phi/2}\widetilde{N}_{i}^{*}),
\label{eq:mass_eigenstates}
\end{eqnarray}
where $\Phi\equiv\arg(BM)$ and with mass eigenvalues
\begin{eqnarray}
M_{ii\pm}^{2} & = & M_{ii}^{2}+\tilde{m}_{ii}^{2}
\pm|B_{ii}M_{ii}|.
\label{eq:mass_eigenvalues}
\end{eqnarray}
In what follows, we will consider a single generation of $N$ 
and $\widetilde{N}$ which we label as $1$. 
We also assume proportionality of soft trilinear terms and drop the
flavour indices for the coefficients $A$ and $B$.  
As discussed in Refs.~\cite{soft1,soft2}, in this case,  after superfield 
rotations the Lagrangians (\ref{eq:superpotential}) and (\ref{eq:soft_terms}) 
have a unique independent physical CP violating phase, 
$\phi={\rm arg}(A B^*)$ which we chose to assign to $A$.

Neglecting supersymmetry breaking effects in the right sneutrino masses
and in the vertex, the total singlet sneutrino decay width is given by  
\be
\Gamma_{\widetilde{N}_+}
=\Gamma_{\widetilde{N}_-}\equiv \Gamma_{\widetilde{N}}
=\frac {\displaystyle M\,  (YY^\dagger)_{11}}{\displaystyle 4 \pi}
\equiv\frac{m_{eff}}{4\pi} \frac{M^2}{v_u^2}  .
\label{eq:gamma}
\end{equation}
where $v_u$ is the vacuum expectation 
value of the up-type Higgs doublet, $v_u=v\, \sin\beta $ ($v$=174 GeV) .

As discussed in Ref.\cite{soft2}, when $\Gamma \gg \Delta
M_{\pm}\equiv M_+ -M_- $, the two singlet sneutrino states are not
well-separated particles. In this case, the result for the asymmetry
depends on how the initial state is prepared. In what follows we will
assume that the sneutrinos are in a thermal bath with a thermalization
time $\Gamma^{-1}$ shorter than the typical oscillation times, $\Delta
M_\pm^{-1}$, therefore coherence is lost and it is appropriate to
compute the CP asymmetry in terms of the mass eigenstates
Eq.(\ref{eq:mass_eigenstates}).

In this regime the relevant BE   
(following ~\cite{kolb,plumacher,ourflasoft} in notation and details)
including the dominant $\Delta L=1,2$ decays and inverse decays 
as well as the $\Delta L=1$ scatterings with top quark are:
\begin{eqnarray}
sHz\frac{dY_{N}}{dz} & =&  
-\left(\frac{Y_{N}}{Y_{N}^{eq}}-1\right)
\left(\gamma_{N}+4\gamma_{t}^{(0)}+4\gamma_{t}^{(1)}
+4\gamma_{t}^{(2)}+2\gamma_{t}^{(3)}+4\gamma_{t}^{(4)}\right),
\label{eq:BEN}
\end{eqnarray}
\begin{eqnarray}
sHz\frac{dY_{\widetilde{N}_{\mbox{tot}}}}{dz} & =  
-\left(\frac{Y_{\widetilde{N}_{\mbox{tot}}}}{Y_{\widetilde{N}}^{eq}}
-2\right)&\left(\gamma_{\widetilde{N}}+\gamma_{\widetilde{N}}^{(3)}
+3\gamma_{22}+2\gamma_{t}^{(5)}+2\gamma_{t}^{(6)}
+2\gamma_{t}^{(7)} \right. \nonumber \\ 
&& \left.
+\gamma_{t}^{(8)}+2\gamma_{t}^{(9)}\right),
\label{eq:BENt} 
\end{eqnarray}
\begin{eqnarray}
sHz\frac{dY_{L_{\mbox{tot}}}}{dz} & = 
\Big[\,\epsilon(T) & \,\left(\frac{Y_{\widetilde{N}_{\mbox{tot}}}}
{Y_{\widetilde{N}}^{eq}}-2\right)-\frac{Y_{L_{\mbox{tot}}}}
{2Y_{c}^{eq}}\Big]\gamma_{\widetilde{N}}
\nonumber \\
 &   -\frac{Y_{L_{\mbox{tot}}}}{2Y_{c}^{eq}}&
\left(\frac{1}{2}\gamma_{N}+\frac{Y_{\widetilde{N}_{\mbox{tot}}}}
{Y_{\widetilde{N}}^{eq}}\gamma_{t}^{(5)}+2\gamma_{t}^{(6)}
+2\gamma_{t}^{(7)}+\frac{Y_{N}}{Y_{N}^{eq}}\gamma_{t}^{(3)}
+2\gamma_{t}^{(4)}\right.\nonumber \\
 &  & \left.
+\gamma_{\widetilde{N}}^{(3)}
+\frac{1}{2}
\frac{Y_{\widetilde{N}_{\mbox{tot}}}}
{Y_{\widetilde{N}}^{eq}}
\gamma_{t}^{(8)}+2\gamma_{t}^{(9)}
+2\frac{Y_{N}}{Y_{N}^{eq}}\gamma_{t}^{(0)}+2\gamma_{t}^{(1)}
+2\gamma_{t}^{(2)}\right)\nonumber \\
 &  -\frac{Y_{L_{\mbox{tot}}}}{2Y_{c}^{eq}}&\left(2+\frac{1}{2}
\frac{Y_{\widetilde{N}_{\mbox{tot}}}}{Y_{\widetilde{N}}^{eq}}\right)
\gamma_{22}.
\label{eq:BE_L_tot}
\end{eqnarray}
In the equations above, $z=M/T$,  
$Y_{\widetilde{N}_{\mbox{tot}}}\equiv 
Y_{\widetilde{N}_{+}}+Y_{\widetilde{N}_{-}}$,  and 
$Y_{L_{\mbox{tot}}} \equiv Y_{L_{f}}+Y_{L_{s}}$ 
with the fermionic and scalar lepton asymmetries 
defined as $Y_{L_f}=(Y_\ell-Y_{\bar\ell})$,
$Y_{L_{s}}=(Y_{\tilde\ell}-Y_{\tilde\ell^*})$. 
The equilibrium abundances are given by
 $Y_{c}^{eq}\equiv\frac{15}{4\pi^{2}g_{s}^{*}}$  
and $Y^{\rm eq}_{\tilde N}(T\gg M) = 90 \zeta(3)/(4\pi^4g_s^*)$,    
where $g_{s}^{*}$ is the total number of entropic degrees of 
freedom, and $g_{s}^{*}=228.75$ in the MSSM. 
In writing Eqs.~(\ref{eq:BEN}--\ref{eq:BE_L_tot}), fast equilibration 
between the lepton asymmetry in scalars and fermions due to supersymmetry
conserving processes has been accounted for.

The different $\gamma$'s are the thermal widths for the following processes:
\begin{eqnarray}
&&\gamma_{\widetilde N}=\gamma^f_{\widetilde N}+\gamma^s_{\widetilde N}=
\gamma(\widetilde{N}_{\pm}\leftrightarrow
\bar{\tilde{h}}\ell)
+\gamma(\widetilde{N}_{\pm} \leftrightarrow h\tilde{\ell}) ,
\nonumber \\
&&\gamma^{(3)}_{\widetilde N}=\gamma( 
\widetilde{N}_{\pm}\leftrightarrow 
\tilde{\ell}^{*}\tilde{u}\tilde{q})\; , 
\nonumber \\
&&\gamma_{22} =  \gamma(\widetilde{N}_{\pm}
\tilde{\ell}\leftrightarrow\tilde{u}\tilde{q}
)=\gamma(\widetilde{N}_{\pm}
\tilde{q}^{*}\leftrightarrow\tilde{\ell}^{*}\tilde{u}
)=\gamma(\widetilde{N}_{\pm}\tilde{u}^{*}\leftrightarrow\tilde{\ell}^{*}\tilde{q}) ,
\nonumber \\
&&\gamma_t^{(5)}=\gamma(\widetilde{N}_{\pm}
\ell\leftrightarrow q\tilde{u})=\gamma(
\widetilde{N}_{\pm}\ell\leftrightarrow\tilde{q}\bar{u})\; ,
\nonumber \\
&&\gamma_t^{(6)}=\gamma(
\widetilde{N}_{\pm}\tilde{u}\leftrightarrow\bar{\ell}q)=\gamma( 
\widetilde{N}_{\pm}\tilde{q}^{*}\leftrightarrow\bar{\ell}\bar{u})\;,
\nonumber \\
&&\gamma_t^{(7)}=\gamma(
\widetilde{N}_{\pm}\bar{q}\leftrightarrow\bar{\ell}\tilde{u})=\gamma( 
\widetilde{N}_{\pm}u\leftrightarrow\bar{\ell}\tilde{q}) , 
\nonumber \\
&&\gamma_t^{(8)}=\gamma(
\widetilde{N}_{\pm}\tilde{\ell}^{*}\leftrightarrow\bar{q}u) ,
\nonumber \\
&&\gamma_t^{(9)}=\gamma(
\widetilde{N}_{\pm}q\leftrightarrow\tilde{\ell}u)= 
\gamma(\widetilde{N}_{\pm}\bar{u}\leftrightarrow\tilde{\ell}\bar{q}) ,
\nonumber \\
&&\gamma_N=\gamma(N\leftrightarrow \ell h)+
\gamma(N\leftrightarrow \tilde{\ell}^* \tilde h) ,
\nonumber \\
&&\gamma_t^{(0)}=\gamma(N\tilde{\ell}\leftrightarrow q\tilde{u})=
\gamma(N\tilde{\ell}\leftrightarrow\tilde{q}\bar{u}) ,
\nonumber \\
&&\gamma_t^{(1)}=\gamma(N\bar{q}\leftrightarrow\tilde{\ell}^{*}\tilde{u})=
\gamma(N\leftrightarrow\tilde{\ell}^{*}\tilde{q})\; ,
\nonumber \\
&&\gamma_t^{(2)}=\gamma(N\tilde{u}^{*}\leftrightarrow\tilde{\ell}^{*}q)=
\gamma(N\tilde{q}^{*}\leftrightarrow\tilde{\ell}^{*}\bar{u})\; , 
\nonumber \\
&&\gamma_t^{(3)}=\gamma(N\ell\leftrightarrow q\bar{u})\; ,
\nonumber \\
&&\gamma_t^{(4)}=\gamma (N\leftrightarrow\bar{\ell}q)=
\gamma(N\bar{q}\leftrightarrow\bar{\ell}\bar{u})\; ,
\label{eq:gammas}
\end{eqnarray}
where in all cases a sum over the CP conjugate final states is implicit.

The final amount of ${B}-{L}$ asymmetry generated by the decay of the 
singlet sneutrino states assuming no pre-existing
asymmetry can be parameterized as:
\begin{equation}
Y_{B-L}(z\rightarrow \infty)=-Y_{L_{\rm tot}}(z\rightarrow \infty)=
-2 \eta\, \bar\epsilon \, Y^{\rm eq}_{\tilde N}(T>>M), 
\label{eq:yb-l}
\end{equation}
where $\bar\epsilon$ is given in Eq.(\ref{eq:epsilon_f_T0})~\footnote{
The factor 2 in Eq.~(\ref{eq:yb-l}) arises from the fact that there are two 
right-handed sneutrino states while we have defined 
$Y^{\rm eq}_{\tilde N}$ for one degree of freedom. Defined this way,
$\eta$ has the standard normalization $\eta\rightarrow 1$ for perfect
out of equilibrium decay.}. 

$\eta$ is a dilution factor which
takes into account the possible inefficiency in the production of the
singlet sneutrinos, the erasure of the generated asymmetry by
$L$-violating scattering processes and the temperature dependence of the 
CP asymmetry  and it is obtained by solving the array
of BE above. 

After conversion by the sphaleron transitions, the final baryon asymmetry
is related to the ${B}-{ L}$ asymmetry by~\cite{sphal}
\begin{equation}
Y_{B}=\frac{8}{23} \, Y_{B-L}(z\rightarrow \infty)\; .
\end{equation}

Without including quantum effects in the BE, the relevant CP asymmetry in 
Eq.~(\ref{eq:BE_L_tot})  is:
\be
\label{epi}
\epsilon (T)\,\equiv\,\bar\epsilon(T) 
\, =\, \frac{\displaystyle \sum_{a_k,k} \gamma(\widetilde{N}_i \rightarrow a_k)
- \gamma(\widetilde{N}_i \rightarrow \bar{a}_k)}
{\displaystyle \sum_{a_k,k} \gamma(\widetilde{N}_i \rightarrow a_k)
+ \gamma(\widetilde{N}_i \rightarrow \bar{a}_k)} \; , 
\end{equation}
where $a_k\equiv s_k,f_k$ with $s_k=\tilde{\ell}_k h$ and 
$f_k=\ell_k \tilde h$ and we denote by $\gamma$ the thermal averaged rates.  

Neglecting supersymmetry breaking in vertices: 
\be
\epsilon (T) \,= \,\bar \epsilon\, \ \frac{c_{s}(T) - c_{f}(T)}{c_{s}(T) + c_{f}(T)} 
\,\equiv\, \bar\epsilon\;\times\; \Delta_{BF} (T)\ ,
 \,
\label{eq:asymunf}
\ee
where 
\begin{equation}
\bar\epsilon\, =\, \frac{{\rm Im}A}{M} 
\frac{4\Gamma B}{4B^{2}+\Gamma^{2}},
\label{eq:epsilon_f_T0}
\end{equation}
is the zero temperature CP asymmetry arising from the scalar (or minus 
the fermion) loop contribution to the $\widetilde N$ self-energy.
The thermal factors are:
\bea
c_f(T)
&=&(1-x_{\ell} -x_{\tilde{h}})\lambda(1,x_{\ell},x_{\tilde{h}})
\left[ 1-\fLeq\right] \left[ 1-\fht\right], 
\label{cfeq}\\
c_s(T)&=&\lambda(1,x_h,x_{\tilde{\ell}})
\left[ 1+\fh\right] \left[ 1+\fLteq\right],
\label{cbeq}
\eea
where
\bea
f^{eq}_{h,\tilde{\ell}}=\frac{1}{\exp[E_{h,\tilde{\ell}}/T]-1}
&\;\;\;\;\;\;\;\;\;\;\;\;&,
f^{eq}_{\tilde h,\ell}= \frac{1}{\exp[E_{\tilde h,\ell}/T]+1}  
\label{eq:fheq},
\eea
are the  Bose-Einstein and Fermi-Dirac equilibrium distributions,
respectively, and 
\bea
&E_{\ell,\tilde h}=\frac{M}{2} (1+x_{\ell,\tilde{h}}-
x_{\tilde h,\ell}), ~~~
E_{h,\tilde{\ell}}=\frac{M}{2} (1+x_{h ,\tilde{\ell}}-
x_{\tilde{\ell},h})&\\
&\lambda(1,x,y)=\sqrt{(1+x-y)^2-4x},~~~
x_a\equiv \frac{m_a(T)^2}{M^2}&.
\eea

We remind the reader that, as seen above, the relevant CP asymmetry
in soft leptogenesis is $T$ (i.e. time) dependent even in the 
classical regime. This is so because the CP asymmetry is generated 
by the supersymmetry breaking thermal effects which make the relevant 
decay CP asymmetries into scalars and fermions different. In the
absence of these thermal corrections, no asymmetry is generated.
 
The inclusion of quantum effects (for technical details see 
Refs.~\cite{riottoqbe1,riottoqbe2,riottoqbe3}) introduces an additional 
time dependence in the CP asymmetry:
\begin{equation}
\epsilon(T)=\epsilon\, \times\,\Delta_{BF}(T)\, \times\, QC(t), 
\end{equation}
where

\begin{equation} 
QC(t) =\left[2 ~\sin^2\left(\frac{M_+-M_-}{2} t\right)
- \frac{\Gamma_{\tilde N}}{M_+-M_-}~\sin\left((M_+-M_-)t\right)\right].
\label{eq:qct}
\end{equation}

Now, we simply have to change the variable from time $t$ of Eq. (\ref{eq:qct}) to a more convenient variable $z$ as we do when writing down the BE as in Eqs. (\ref{eq:BEN} - \ref{eq:BE_L_tot}). As a reminder to the readers, we will write a few lines illustrating this change.
For a universe undergoing adiabatic expansion, the entropy per comoving volume is constant i.e. 
$sR^3 =$ constant. Since $s \propto z^{-3}$, we have $R \propto z$. 
Then, the Hubble constant is given by $H \equiv R^{-1}dR/dt=z^{-1}dz/dt$. 
After integration, we get

\begin{equation}
t=\frac{1}{H(M)}\frac{z^2-z_0^2}{2},
\label{eq:t_to_z}
\end{equation}
where $z_0$ is the temperature at $t=0$, and $H(M)\equiv H(z=1)=2/3 \sqrt{g^*\pi^3/5}(M^2/m_{\rm pl})$ with 
$m_{\rm pl}$ being the Planck mass. Substituting Eq. (\ref{eq:t_to_z}) into Eq. (\ref{eq:qct}), 
we obtain 

\begin{eqnarray} 
QC(T) &=&\left[2 ~\sin^2\left(\frac{1}{2} \frac{M_+-M_-}{2 H(M)} z^2\right)
- \frac{\Gamma_{\tilde N}}{M_+-M_-}~\sin\left(\frac{M_+-M_-}{2 H(M)}z^2\right)\right],\nonumber \\
&=&\left[2\sin^2\left(\frac{m_{eff}}{m_*}\, R\, \frac{z^2}{8}\right)-
\frac{2}{R}\sin\left(\frac{m_{eff}}{m^*}\, R\, \frac{z^2}{4}\right)\right].
\label{eq:qc}
\end{eqnarray}
where we set $z_0 = 0$ (i.e. at very high initial temperature). In writing the second equality 
we have used that $M_+-M_-=B$  
(see Eq.~(\ref{eq:mass_eigenvalues})), and we have defined the 
degeneracy parameter $R$, 
\begin{equation}
R= \frac{2(M_+-M_-)}{\Gamma_{\widetilde N}}=
\frac{2B}{\Gamma_{\widetilde N}}, 
\end{equation}
and
\begin{equation}
m_*=\frac{8 v_u^2}{3 m_{\rm pl}}\sqrt{\frac{g_* \pi^5}{5}}
\simeq 7.8\times 10^{-4}\; {\rm eV}.
\end{equation}

Thus the final CP asymmetry consists of three factors. The first one is 
$\bar\epsilon$ in Eq.~(\ref{eq:epsilon_f_T0}) 
\begin{equation}
\bar\epsilon = \frac{{\rm Im}A}{M} \frac{2R}{R+1},
\end{equation}
which is resonantly enhanced for $R=1$. The second one is the thermal
factor $\Delta_{BF}(T)$ which is only non-vanishing for $z\gsim 0.8$
 \cite{soft3,thermal}.
The third one is the quantum correction factor, $QC(T)$ 
which is composed of two oscillating functions.    

Next we turn to quantify the impact of this last additional quantum 
time-dependence of the  CP asymmetry on the final lepton asymmetry.

Before doing so, let us notice that Eq.~(\ref{eq:BE_L_tot})  
corresponds to the one-flavour approximation. 
As discussed in Refs.~\cite{flavour1,flavour2,oscar,db1,PU,
riottoqbe3,barbieri,antusch,riottosc,flavourothers} 
the one-flavour approximation
is rigorously correct only when the interactions mediated by charged
lepton Yukawa couplings are out of equilibrium.  This is not the case
in soft leptogenesis since successful
leptogenesis in this scenario requires a relatively low RH
neutrino mass scale.
Thus the characteristic $T$ is such that the
rates of processes mediated by the $\tau$ and $\mu$ Yukawa couplings
are not negligible implying that the effects of lepton flavours have
to be taken into account~\cite{ourflasoft}. 

However, as shown in Refs.~\cite{riottoqbe1,riottoqbe2,riottoqbe3} 
quantum effects are flavour independent as long as 
the damping rates of the leptons  are taken to be flavour independent.
In this case the $QC(T)$ factor becomes the one given above (neglecting also
the difference in the width between the two sneutrinos) which is 
the same for all flavours. Furthermore 
quantum flavour correlations can be safely neglected for 
soft leptogenesis because $M\lsim 10^{-9}$ GeV and therefore 
there is no transition between three-to-two or two-to-one flavour regimes. 
So following  Ref.~\cite{ourflasoft}, 
it is straight forward to include flavour in the QBE given above 
in terms of flavour dependent wash-out factors. 
For the sake of simplicity in the presentation we restrict here
to the one-flavour approximation for the study of the relevance
of the quantum effects.  

\section{Results}
\begin{figure}
\begin{center}
\hspace*{2.5cm}
\includegraphics[width=0.9\textwidth]{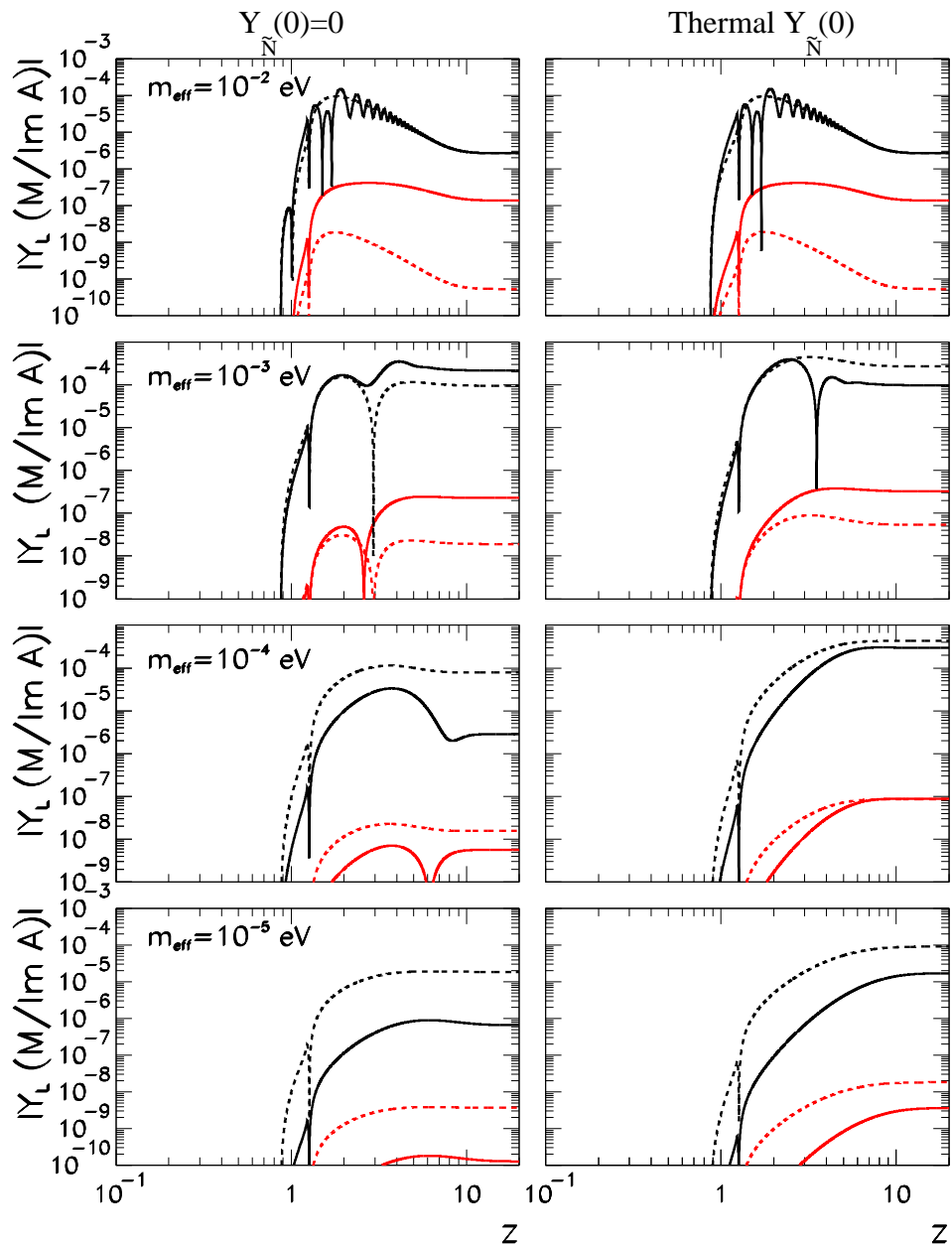}
\caption{Absolute value of the lepton asymmetry with the quantum
time dependence of the CP asymmetry (solid) and without it (dashed)
as a function of $z$  for different values of $m_{eff}$ as 
labeled in the figure. 
In each panel the two upper curves (black) correspond to the 
resonant case $R=1$ while the lower two curves (red) correspond to the 
very degenerate case $R=2\times 10^{-4}$. 
The left (right) panels correspond 
to vanishing (thermal) initial $\tilde N$ abundance.
The figure is shown for $M=10^7$ GeV and $\tan\beta=30$ though as discussed
in the text, the results as normalized in the figure are very weakly dependent 
on those two parameters.}
\label{fig:effz}
\end{center}
\end{figure}

\begin{figure}
\begin{center}
\hspace*{2.5cm}
\includegraphics[width=0.9\textwidth]{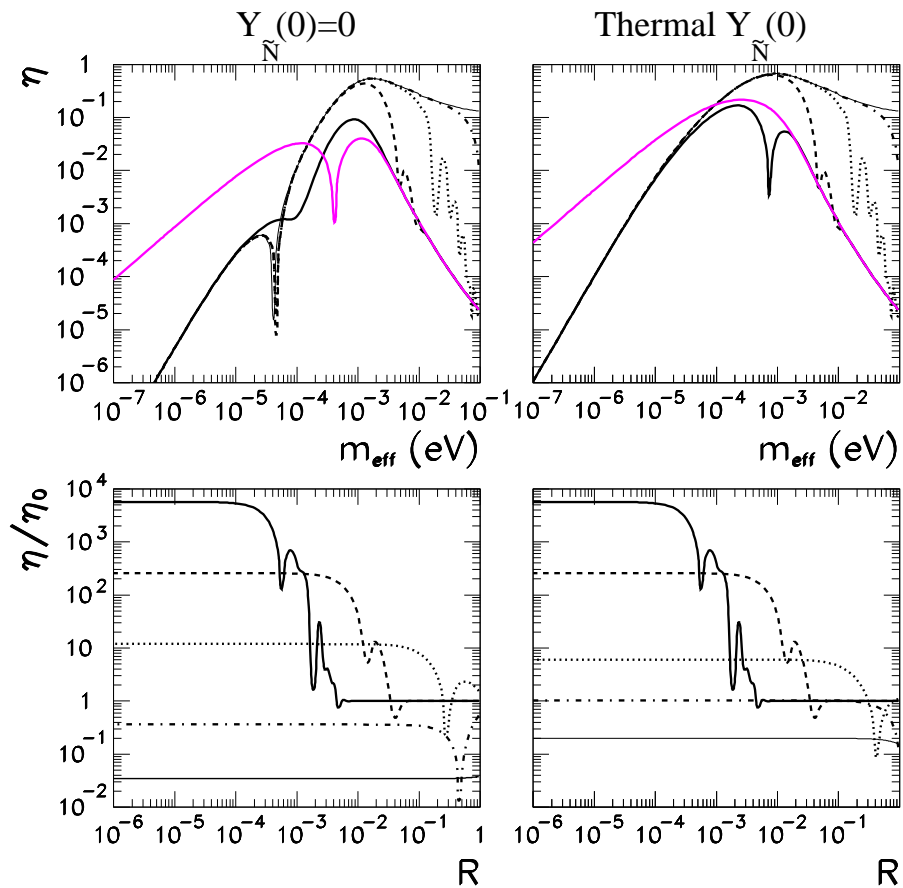}
\caption{Efficiency factor as a function of $m_{eff}$ and $R$ 
for $M=10^{7}$ GeV and $\tan\beta=30$.
The left (right) panels 
correspond to  vanishing (thermal) initial $\tilde N$ abundance.
In the upper panels the different curves correspond to 
$R=1$ (black thick solid) , 0.1 (dashed), $10^{-2}$ (dotted),
$10^{-3}$ (dash-dotted) and $10^{-4}$ (thin solid).
For comparison we also show the results without including the quantum
effects (purple thick solid line).  
In the lower panels we plot the ratio of the efficiency factor with 
and without quantum corrections as a function of $R$.
The different curves from top to bottom
correspond to $m_{eff}=10^{-1}$ eV (think solid), $10^{-2}$ eV (dashed),
$10^{-3}$ eV (dotted), $10^{-4}$ eV (dot-dashed), and 
$10^{-5}$ eV (thin solid).}  
\label{fig:eff}
\end{center}
\end{figure}

\begin{figure}
\begin{center}
\hspace*{2.5cm}
\includegraphics[width=0.9\textwidth]{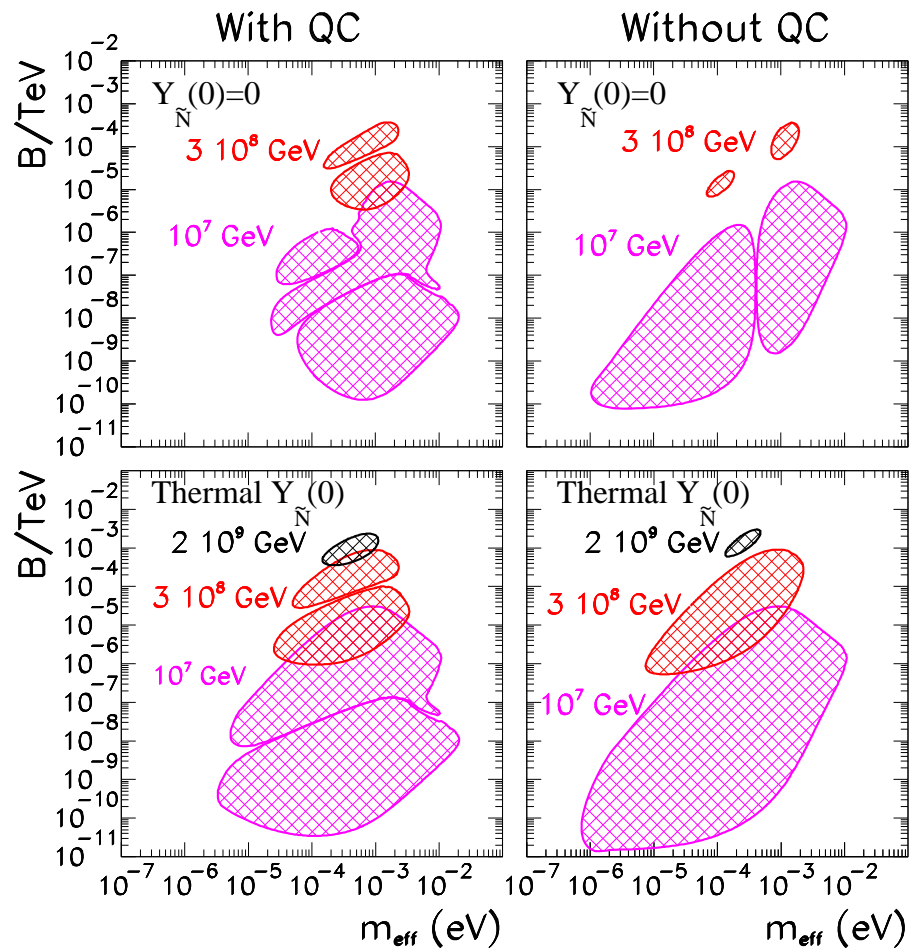}
\caption{
$B, m_{eff}$ regions in which successful soft leptogenesis can
be achieved when quantum effects are with (left panels) 
and without (right panels) quantum effects. 
We take $|{\rm Im} A|=10^3$ GeV and $\tan\beta=30$ and different
values of $M$  as labeled in the figure.
The upper (lower) panels correspond to  
vanishing (thermal) initial 
$\tilde N$ abundance.}
\label{fig:con}
\end{center}
\end{figure}

We show in Fig.~\ref{fig:effz} the evolution of the lepton asymmetry
with and without the quantum correction factor in the CP asymmetry for
several values of the washout factor $m_{eff}$ and for the resonant
case $R=1$ and the very degenerate case $R=2\times 10^{-4}$. The two
upper panels correspond to strong and moderate washout regimes, while the
lower two correspond to weak and very weak washout regimes.  We
consider two different initial conditions for the sneutrino abundance.
In one case, one assumes that the ${\tilde N}$ population is created
by their Yukawa interactions with the thermal plasma, and set
$Y_{\tilde N}(z\rightarrow 0)=0$. The other case corresponds to an
initial $\tilde N$ abundance equal to the thermal one, 
$Y_{\tilde
N}(z\rightarrow 0)= Y_{\tilde N}^{eq}(z\to 0)$.  
The initial condition on the sneutrino abundances can lead to differences in 
the weak washout regime. In the strong washout regime the asymmetry
generated in the $\widetilde N$ production phase for 
$Y_{\tilde N}(z\rightarrow 0)= Y_{\tilde N}^{eq}(z\to 0)$  is 
efficiently washed out (contrary to what happens in the weak 
washout regime). Consequently, in the strong washout regime the generated
asymmetry is independent of the initial conditions. This behaviour is
explicitly displayed on the upper panel of Fig.~\ref{fig:effz}.
It can also be observed   on the right hand side of the upper panels 
as well as on the  upper curves 
of the lower panels of Fig.~\ref{fig:eff}, and on the right hand side of 
Fig.~\ref{fig:con}.

First we notice that, as expected, for strong washout and large 
degeneracy parameter $R$ (see the upper curves in the upper panels), 
the quantum effects lead to the oscillation of the produced asymmetry 
till it finally averages out to the {\sl classical} value.  

The figure also illustrates that for very small value of $R$ and 
in the strong washout regime,  quantum effects enhance the final asymmetry. 
For small enough $R$ the arguments in the periodic functions in 
$QC(T)$  are very small for all relevant values of $z$ and $m_{eff}$. 
So the $\sin^2$ term in  $QC(T)$ is negligible and expanding  
the $\sin$ term  we get 
\begin{equation}
QC(T)\simeq - \frac{m_{eff}}{m_*}\frac{z^2}{2},  
\label{eq:qclim}
\end{equation}
which, in the strong washout regime is always larger than 1. 

Also we see that, independently of the initial conditions,
and of the value of the degeneracy parameter, $R$, the quantum effects 
always lead to a suppression
of the final produced asymmetry in the weak washout regime. 
This is at difference of what happens in see-saw resonant leptogenesis
in which quantum effects lead to an enhancement of the produced asymmetry
in  weak washout and $R\sim 1$ and for zero initial
sneutrino abundances~\cite{riottoqbe2}
~\footnote{We notice in passing that for standard see-saw resonant 
leptogenesis the weak washout regime is physically unreachable 
as long as flavour effects are not included. This is so because
there is a lower bound on the washout parameter once the washout 
associated to the two quasi-degenerate heavy neutrinos contributes 
which implies that 
$\tilde m\geq \sqrt{\Delta m^2_{\rm solar}}\sim 8\times 10^{-3}$ 
\cite{dibari2}. 
Such bound does not apply to soft leptogenesis as long as, as assumed 
in this work, only the lightest sneutrino generation contributes.}.  

The origin of the difference 
is the additional time dependence of the asymmetry in soft leptogenesis
due to $\Delta_{BF}$. In order to understand this, we 
must remember that  in see-saw resonant leptogenesis, in the weak
washout regime, the final lepton asymmetry results from a cancellation
between the anti-asymmetry generated when RH neutrinos are
initially produced and the lepton asymmetry produced when they finally 
decay. When the time-dependent quantum corrections are included,  
this near-cancellation does not hold or it occurs at earlier times. 
As a consequence the asymmetry grows larger once these corrections are 
included as discussed in Ref.~\cite{riottoqbe2}. 

But in soft-leptogenesis,  even in the classical regime  
the thermal factor $\Delta_{BF}$  already prevents the cancellation 
to occur. Therefore the inclusion of the time dependent quantum 
effects only amounts to an additional multiplicative 
factor which, in this regime, is smaller than one. 

\begin{figure}
\begin{center}
\hspace*{2.5cm}
\includegraphics[width=0.9\textwidth]{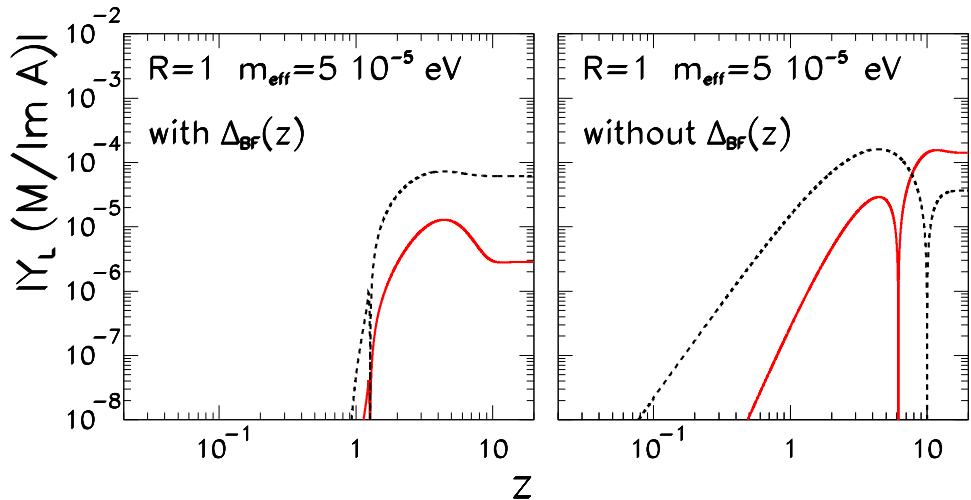}
\caption{
Absolute value of the lepton asymmetry with the quantum
time dependence of the CP asymmetry (solid) and without it (dashed)
for  vanishing initial $\tilde N$ abundance. For comparison 
in the right panel we show the result that would be obtained with 
$\Delta_{BF}(z)=1$.}   
\label{fig:effzt}
\end{center}
\end{figure}

This behaviour is explicitly displayed in Fig.~\ref{fig:effzt} 
where we compare the absolute value of the lepton asymmetry with the quantum
time dependence of the CP asymmetry and without it in soft
leptogenesis with what would be obtained if the thermal factor  
$\Delta_{BF}(z)$ was not included (so that the CP asymmetry takes 
a form similar to the one for resonant see-saw).  
As seen in Fig.~\ref{fig:effzt}, without the $\Delta_{BF}(z)$ 
the asymmetry starts being produced at lower $z$ and 
it changes sign in the classical regime. This change
of sign is due to the cancellation  
between the anti-asymmetry generated when RH neutrinos are
initially produced and the lepton asymmetry produced when they finally 
decay. Inclusion of the $QC(T)$ factor reduces the asymmetry 
at small $z$ and this makes the cancellation to occur at lower 
$z$ and consequently the final asymmetry is larger.

In the full calculation (left panel in Fig.~\ref{fig:effzt}) 
the asymmetry only starts
being non-negligible for larger $z$, i.e. $z\gsim 0.8$, 
and it changes sign for $z\sim 1$ , both features due to 
the $\Delta_{BF}$ factor. Inclusion of the quantum correction, 
$QC(T)$ amounts for a suppression of the initial asymmetry 
by a factor given in Eq.~(\ref{eq:qclim}).
As a consequence the final asymmetry is suppressed (and it also has
the opposite sign) after including the quantum corrections.

A more systematic dependence of the results with the 
washout and degeneracy parameters, $m_{eff}$ and $R$ 
is shown in  Fig.~\ref{fig:eff} where we plot the efficiency 
factor $\eta$ as a function of  $m_{eff}$ and $R$. 
We remind the reader that within our approximations for the 
thermal widths, in the classical regime, $\eta$ is mostly 
a function of $m_{eff}$ exclusively~\footnote{There is a residual dependence 
on $M$ due to the running of 
the top Yukawa coupling as well as the thermal effects included in 
$\Delta_{BF}$ although it is very mild.
For
$\tan\beta\sim {\cal O}(1)$ there is also an additional (very weak)
dependence due to the associated change in the top Yukawa coupling.
}. Inclusion of the 
quantum correction $QC(T)$ makes $\eta$ to depend both on
$m_{eff}$ and $R$ but still remains basically independent of $M$.

From the figure we see that for small enough values of the product
of the washout parameter and the degeneracy parameter  the arguments
of the periodic functions in $QC(T)$ are always small in the range
of $z$ where the lepton asymmetry is generated. As explained above, 
in this regime the 
$\sin^2$ term in  $QC(T)$ is negligible while the $\sin$ term is multiplied 
by an amplitude proportional to $1/R$. Therefore, the
dependence on $R$ cancels in this limit and the resulting correction is
given in Eq.~(\ref{eq:qclim}). This explains the plateaux observed
at low values of the degeneracy parameter $R$ in the lower panels of 
Fig.~\ref{fig:eff}. Similar behaviour is found in Ref.\cite{riottoqbe3}
for the resonant leptogenesis scenario.
Also, as seen in Eq.~(\ref{eq:qclim}), the correction
grows with $m_{eff}$ which leads to the considerable enhancement of the 
efficiency seen in the upper curves of the lower panel in Fig.~\ref{fig:eff}. 
However we must notice that this enhancement occurs in a
regime where the CP asymmetry is very small due to the small value
of $R$ since $\bar\epsilon$ is proportional to $R$.

Finally, in  Fig. \ref{fig:con} we compare the range of parameters
$B$ and $m_{eff}$ for which enough asymmetry is
generated, $Y_B\geq 8.54\times 10^{-11}$ with and without inclusion 
of the quantum corrections.
We show the ranges for several values of $M$  and for the characteristic 
value of  $|{\rm Im} A|=1$ TeV.  From the figure we see that due to the
suppression of the asymmetry for the weak washout regime discussed above, for
a given value of $M$ the regions extend only up to larger values of 
$m_{eff}$ once the quantum corrections are included. Also, because of the
enhancement in the very degenerate, strong washout regime, the regions tend
to extend to lower values of $B$ and larger values of $m_{eff}$ for a given
value of $M$. Furthermore, once quantum effects are included,  
$\eta$ can take both signs  (depending on the value of $m_{eff}$), 
independently of the initial $\tilde N$ abundance.
Thus it is possible to generate the right
sign asymmetry with either sign of ${\rm Im} A $
for both thermal and zero initial $\tilde N$ abundance.  On the contrary
without quantum corrections, for thermal initial conditions $\eta>0$ 
and the right asymmetry can only be generated for ${\rm Im} A>0 $.  

\section{Summary}
In this article  we have performed a detailed study 
of the role of quantum effects in the soft leptogenesis scenario. 
We have studied the effects on the produced asymmetry as a function 
of the washout parameter $m_{eff}$ and the degeneracy parameter
$R=2\Delta M/\Gamma_{\widetilde N}$
Our results show that, because of the thermal nature of soft supersymmetry, 
the characteristic time for the building of the asymmetry is larger than 
in the see-saw resonant leptogenesis which leads to quantitative differences
on the dependence of the effect on the washout regime between the two 
scenarios.  In particular, in the weak washout
regime, quantum effects do not enhance but suppress the produced 
lepton asymmetry in soft leptogenesis.  
Quantum effects are most quantitatively important 
for extremely degenerate sneutrinos  
$\Delta M \ll \Gamma_{\widetilde N}$. In this case and in the strong washout 
regime quantum effects can enhance the absolute value of the 
produced asymmetry as well as induce a change of its sign.  
But altogether, our results show that the required 
values of the Majorana mass $M$ and the lepton violating soft bilinear 
coefficient $B$ to achieve successful leptogenesis are not 
substantially modified. 

\section*{acknowledgments} 
This work is supported by National Science Foundation
grant PHY-0354776 and  by Spanish Grants FPA-2004-00996
and  FPA2006-28443-E.


\vskip 1cm
\begin{thebibliography}{99}
%
\bibitem{fy}
M. Fukugita and T. Yanagida, Phys. Lett. {\bf B174} (1986) 45
%
\bibitem{leptoreview}
 S.~Davidson, E.~Nardi and Y.~Nir,
  arXiv:0802.2962 [hep-ph].
%
\bibitem{ss} 
P.~Minkowski,
  Phys.\ Lett.\ B {\bf 67}, 421 (1977);
M. Gell-Mann, P. Ramond and R. Slansky, Proceedings of   
the Supergravity Stony Brook Workshop, New York, 1979, eds. P. Van   
Nieuwenhuizen and D. Freedman (North-Holland, Amsterdam);
T. Yanagida, Proceedings of
the  Workshop  on Unified  Theories  and  Baryon  Number in the  
Universe,  Tsukuba,  Japan 1979 (eds. A.  Sawada and A.
Sugamoto, KEK Report No.  79-18, Tsukuba); 
R.~Mohapatra and G.~Senjanovic, 
Phys.\ Rev.\ Lett.\ {\bf 44}, 912 (1980).
%
\bibitem{di}
  S.~Davidson and A.~Ibarra,
  Phys.\ Lett.\ B {\bf 535} (2002) 25
  [arXiv:hep-ph/0202239].
%
\bibitem{Mbound}
W.~Buchmuller, P.~Di Bari and M.~Plumacher,
Nucl.\ Phys.\ B {\bf 643} (2002) 367 [arXiv:hep-ph/0205349];
J.~R.~Ellis and M.~Raidal,
Nucl.\ Phys.\ B {\bf 643} (2002) 229 [arXiv:hep-ph/0206174].
%
\bibitem{flavour1}
 A.~Abada, S.~Davidson, A.~Ibarra, F.~X.~Josse-Michaux, M.~Losada and 
 A.~Riotto,  
  arXiv:hep-ph/0605281;
  A.~Abada, S.~Davidson, F.~X.~Josse-Michaux, M.~Losada and A.~Riotto,
  JCAP {\bf 0604} (2006) 004
  [arXiv:hep-ph/0601083];
%
\bibitem{flavour2}
 E.~Nardi, Y.~Nir, E.~Roulet and J.~Racker,
  JHEP {\bf 0601}, 164 (2006)
  [arXiv:hep-ph/0601084];
%
\bibitem{db2}
 S.~Blanchet and P.~Di Bari,
  arXiv:hep-ph/0607330.
%
\bibitem{oscar}
 O.~Vives,
  Phys.\ Rev.\ D {\bf 73} (2006) 073006
  [arXiv:hep-ph/0512160].
%
\bibitem{ma}
 E.~Ma, N.~Sahu and U.~Sarkar,
  J.\ Phys.\ G {\bf 32}, L65 (2006)

\bibitem{db1}
 P.~Di Bari,
  Nucl.\ Phys.\ B {\bf 727} (2005) 318
  [arXiv:hep-ph/0502082].
%
\bibitem{gravi} 
M.~Y.~Khlopov and A.~D.~Linde,
Phys.\ Lett.\ B \textbf{138} (1984) 265;
J.~R.~Ellis, J.~E.~Kim and D.~V.~Nanopoulos,
Phys.\ Lett.\ B \textbf{145} (1984) 181;
J.~R.~Ellis, D.~V.~Nanopoulos and S.~Sarkar,
Nucl.\ Phys.\ B \textbf{259} (1985) 175;
T.~Moroi, H.~Murayama and M.~Yamaguchi,
Phys.\ Lett.\ B \textbf{303} (1993) 289;
M.~Kawasaki, K.~Kohri and T.~Moroi,
Phys.\ Lett.\ B {\bf 625} (2005) 7;
For a recent discussion, see:
  K.~Kohri, T.~Moroi and A.~Yotsuyanagi,
  Phys.\ Rev.\ D {\bf 73} (2006) 123511
%
%
\bibitem{PU1}
  A.~Pilaftsis and T.~E.~J.~Underwood,
  Nucl.\ Phys.\  B {\bf 692} (2004) 303
  [arXiv:hep-ph/0309342].

\bibitem{PU}
A.~Pilaftsis and T.~E.~J.~Underwood,
 Phys.\ Rev.\ D {\bf 72} (2005) 113001
  [arXiv:hep-ph/0506107].
 A.~Anisimov, A.~Broncano and M.~Plumacher,
  Nucl.\ Phys.\ B {\bf 737} (2006) 176
  [arXiv:hep-ph/0511248].
%
\bibitem{west}
 T.~Hambye, J.~March-Russell and S.~M.~West,
  JHEP {\bf 0407} (2004) 070
  [arXiv:hep-ph/0403183];
 S.~M.~West,
  Mod.\ Phys.\ Lett.\  A {\bf 21} (2006) 1629;
%
\bibitem{soft1}
Y.~Grossman, T.~Kashti, Y.~Nir and E.~Roulet,
  Phys.\ Rev.\ Lett.\  {\bf 91} (2003) 251801
  [arXiv:hep-ph/0307081];
%
\bibitem{soft2}
  G.~D'Ambrosio, G.~F.~Giudice and M.~Raidal,
  Phys.\ Lett.\ B {\bf 575}, 75 (2003)
  [arXiv:hep-ph/0308031].
%
\bibitem{ourflasoft}  
C.~S.~Fong and M.~C.~Gonzalez-Garcia,
  arXiv:0804.4471 [hep-ph].
%
\bibitem{soft3}
  Y.~Grossman, T.~Kashti, Y.~Nir and E.~Roulet,
  JHEP {\bf 0411} (2004) 080
  [arXiv:hep-ph/0407063].
%
\bibitem{ourinvsoft}
 J.~Garayoa, M.~C.~Gonzalez-Garcia and N.~Rius,
  JHEP {\bf 0702} (2007) 021
  [arXiv:hep-ph/0611311].
%
\bibitem{softothers}
G.~D'Ambrosio, T.~Hambye, A.~Hektor, M.~Raidal and A.~Rossi,
  Phys.\ Lett.\ B {\bf 604} (2004) 199
  [arXiv:hep-ph/0407312];
  M.~C.~Chen and K.~T.~Mahanthappa,
  Phys.\ Rev.\  D {\bf 70}, 113013 (2004)
  [arXiv:hep-ph/0409096];
 Y.~Grossman, R.~Kitano and H.~Murayama,
  JHEP {\bf 0506}, 058 (2005)
  [arXiv:hep-ph/0504160];
 E.~J.~Chun and S.~Scopel,
  Phys.\ Lett.\  B {\bf 636}, 278 (2006)
  [arXiv:hep-ph/0510170];
A.~D.~Medina and C.~E.~M.~Wagner,
  JHEP {\bf 0612}, 037 (2006)
  [arXiv:hep-ph/0609052];
  E.~J.~Chun and L.~Velasco-Sevilla,
  JHEP {\bf 0708}, 075 (2007)
  [arXiv:hep-ph/0702039].
%
\bibitem{qbebuch}
  W.~Buchmuller and S.~Fredenhagen,
  Phys.\ Lett.\  B {\bf 483}, 217 (2000)
  [arXiv:hep-ph/0004145].
%
\bibitem{riottoqbe1}
  A.~De Simone and A.~Riotto,
  JCAP {\bf 0708} (2007) 002
  [arXiv:hep-ph/0703175]
\bibitem{riottoqbe2}
  A.~De Simone and A.~Riotto,
  JCAP {\bf 0708} (2007) 013
  [arXiv:0705.2183 [hep-ph]].
\bibitem{riottoqbe3}
  V.~Cirigliano, A.~De Simone, G.~Isidori, I.~Masina and A.~Riotto,
  JCAP {\bf 0801} (2008) 004
  [arXiv:0711.0778 [hep-ph]].
%
\bibitem{sphal}
  S.~Y.~Khlebnikov and M.~E.~Shaposhnikov,
  Nucl.\ Phys.\  B {\bf 308} (1988) 885.
%
\bibitem{kolb}
E.~W.~Kolb and S.~Wolfram,
  Nucl.\ Phys.\ B {\bf 172}, 224 (1980)
  [Erratum-ibid.\ B {\bf 195}, 542 (1982)].
\bibitem{plumacher}
M. Pl\"{u}macher, Nucl. Phys. B\textbf{ 530} 207-246(1998) 
[arXiv:hep-ph/9704231].

\bibitem{barbieri}
R.~Barbieri, P.~Creminelli, A.~Strumia and N.~Tetradis,
Nucl.\ Phys.\ B {\bf 575} (2000) 61
[arXiv:hep-ph/9911315].
%
\bibitem{antusch}S. Antusch, S. F. King, and A. Riotto, 
JCAP \textbf{0611}
[arViv:hep-ph/0609038].
%
\bibitem{riottosc}
 A.~De Simone and A.~Riotto,
 JCAP {\bf 0702} (2007) 005
 [arXiv:hep-ph/0611357].
%
%
\bibitem{flavourothers}
T.~Endoh, T.~Morozumi and Z.~h.~Xiong,
  Prog.\ Theor.\ Phys.\  {\bf 111}, 123 (2004)
[arXiv:hep-ph/0308276]; 
T.~Fujihara, S.~Kaneko, S.~Kang, D.~Kimura, T.~Morozumi and M.~Tanimoto,
  Phys.\ Rev.\ D {\bf 72}, 016006 (2005)
[arXiv:hep-ph/0505076];
S.~Pascoli, S.~T.~Petcov and A.~Riotto,
arXiv:hep-ph/0609125;
  G.~C.~Branco, R.~Gonzalez Felipe and F.~R.~Joaquim,
  Phys.\ Lett.\  B {\bf 645} (2007) 432
  [arXiv:hep-ph/0609297];
S.~Antusch and A.~M.~Teixeira,
arXiv:hep-ph/0611232;
S.~Pascoli, S.~T.~Petcov and A.~Riotto,
arXiv:hep-ph/0611338;
S.~Blanchet, P.~Di Bari and G.~G.~Raffelt,
arXiv:hep-ph/0611337.
%
\bibitem{thermal}
 G.~F.~Giudice, A.~Notari, M.~Raidal, A.~Riotto and A.~Strumia,
  Nucl.\ Phys.\ B {\bf 685} (2004) 89
  [arXiv:hep-ph/0310123].
\bibitem{dibari2}
S.~Blanchet and P.~Di Bari,
  JCAP {\bf 0606}, 023 (2006)
  [arXiv:hep-ph/0603107].
\end{thebibliography}
\end{document}